\documentclass[onecolumn, doublespace, 12pt fonts]{aastex6}
\usepackage{graphicx,latexsym,amssymb, epsfig}
\usepackage{multirow,amsmath,array,booktabs}
\usepackage{subfigure}
\usepackage{tabularx}
\usepackage{color}
\usepackage{bm}
\usepackage[figuresright]{rotating}
\usepackage[section]{placeins}
\usepackage{slashed}
\usepackage{graphicx}
\usepackage{tikz}

\newcommand{\mc}{\mathcal}
\newcommand{\sig}{\sigma}
\newcommand{\gam}{\gamma}

\begin{document}
	\title{The Hadron-Quark Crossover in Neutron Star within Gaussian Process Regression Method}
	\author{Kaixuan Huang\altaffilmark{1}, Jinniu Hu,\altaffilmark{1}, Ying Zhang\altaffilmark{2}, Hong Shen\altaffilmark{1}}
	\altaffiltext{1}{School of Physics, Nankai University, Tianjin 300071,  China} 
	\altaffiltext{2}{Department of Physics, Faculty of Science, Tianjin University, Tianjin 300072, China}
	
	\email{hujinniu@nankai.edu.cn; yzhang@tju.edu.cn;songtc@nankai.edu.cn}

\begin{abstract}
The equations of state of the neutron star at the hadron-quark crossover region are interpolated with the Gaussian process regression (GPR) method, which can reduce the randomness of present interpolation schemes. The relativistic mean-field (RMF) model and Nambu-Jona-Lasinio (NJL) model are employed to describe the hadronic phase and quark phase, respectively. In the RMF model, the coupling term between $\omega$ and $\rho$ mesons is considered to control the density-dependent behaviors of symmetry energy, i.e. the slope of symmetry energy, $L$. Furthermore, the vector interaction between quarks is included in the NJL model to obtain the additional repulsive contributions. Their coupling strengths and the crossover windows are discussed in the present framework under the constraints on the neutron star from gravitational wave detections, massive neutron star measurements, mass-radius simultaneous observation of NICER collaboration, and the neutron skin thickness of $^{208}$Pb from PREX-II. It is found that the slope of symmetry energy, $L$ should be around \replaced{$70\sim 90$ MeV}{$50-90$ MeV} and the crossover window is $(0.3,~0.6)~\rm fm^{-3}$ with these observables. Furthermore, the uncertainties of neutron star masses and radii in the hadron-quark crossover regions are also predicted by the GPR method.		
	\end{abstract}

	
\section{Introduction}
	Great achievements in the observations of the compact star have been obtained recently from two aspects, which inspired many new developments in the relevant subjects. The gravitational-wave (GW) signal from the binary neutron-star (BNS) merger offers a great opportunity to probe the inner structures of the neutron star (NS), which was first detected by the LIGO and Virgo Scientific Collaborations (LVC) in 2017, i.e., the GW170817 event~\citep{Abbott2017a,Abbott2017b}. Its electromagnetic counterpart was rapidly found later on and the era of multi-messenger astronomy was started. The analysis of the GW170817 event extracted the upper limit of tidal effects in BNS, which provides a new observable of NS besides its radius and mass~\citep{Abbott2018}. The two NS radii from GW170817 were estimated to be $R_1=11.9^{+1.4}_{-1.4}$ km and $R_2=11.9^{+1.4}_{-1.4}$ km at the 90\% credible level if the equation of state (EoS) supports the NSs with masses larger than $1.97 M_{\odot}$. Another detection of a compact binary coalescence, GW190814, reported by LVC involves a $22.2-24.3M_{\odot}$ black hole and a compact object with a mass of $2.50-2.67M_{\odot}$~\citep{Abbott2020}. In the absence of additional information, the secondary component of the binary was inconclusive, which might be either the heaviest NS or the lightest black hole ever discovered. Over the past three years, significant efforts have been devoted to studying the secondary object, which could be a rapidly rotating neutron star with exotic degrees of freedom~\citep{Most2020,Li2020,Demircik2021,Dexheimer2021}, the most massive NS~\citep{Huang2020,Tan2020}, the fastest pulsar~\citep{Zhang2020, Zhou2021}, or a binary black hole merger~\citep{Fattoyev2020,Sedrakian2020,Tews2020}.
	
	Another achievement to probe the properties of compact star is the observation of the X-ray emission from several hot spots of NS surface~\citep{Bogdanov2019}. In 2019, Neutron star Interior Composition Explorer (NICER) collaboration reported an accurate measurement of the mass and radius of PSR J0030+0451, a mass of $1.44_{-0.14}^{+0.15}M_\odot$ with a radius of $13.02_{-1.06}^{+1.24}$ km~\citep{Miller2019} and a mass of $1.34_{-0.16}^{+0.15}M_\odot$ with a radius of $12.71_{-1.19}^{+1.14}$ km~\citep{Riley2019} by two independent analysis groups. Recently, the radius of another pulsar, PSR J0740+6620 with mass $2.08^{+0.07}_{-0.07}M_{\odot}$~\citep{Cromartie2020,Fonseca2021}, was reported by two independent groups~\citep{Miller2021,Riley2021} based on NICER and X-ray Multi-Mirror (XMM-Newton) observations. The inferred radius of this massive NS is constrained to  $12.39_{-0.98}^{+1.30}$ km for the mass $2.072^{+0.067}_{-0.066}M_\odot$ by Riley $et~ al.$~\citep{Riley2021} and $13.7_{-1.5}^{+2.6}$ km for the mass $2.08M_\odot$ by Miller $et~al.$~\citep{Miller2021} at 68\% credible level. 
	
	The equation of state (EoS) of the super-dense matter is an essential input and completely indispensable for understanding the processes in heavy-ion collision experiments, core-collapse supernovae, and the properties of NS, like mass, radius, and tidal deformability. The present observations from the NS require that the EoS of nuclear matter must be moderately soft at relatively low densities to obtain a smaller radius and must be stiff enough at high densities to generate a heavier mass. Besides the constraints from the universe, the nuclear experiments in the terrestrial laboratory provided stringent constraints on the EoS around the nuclear saturation densities, such as the saturation density, binding energy per nucleon, incompressibility, and symmetry energy. The density dependence of symmetry energy will determine the behavior of neutron-rich EoS in the high-density region. However, the slope of symmetry energy at nuclear saturation density is still not well determined. Recently, the Lead Radius Experiment(PREX-\uppercase\expandafter{\romannumeral2}) at Jefferson Lab updated the value of the neutron skin thickness of $^{208}{\rm Pb}$ as $R_{\rm skin}^{208}=0.283\pm0.071$ fm~\citep{Adhikari2021}, which was originally reported from the PREX-\uppercase\expandafter{\romannumeral1}~\citep{Abrahamyan2012}. The neutron skin thickness has a strong linear correlation with the slope of symmetry energy, $L$. The strong correlation between $L$ and $R_{\rm skin}^{208}$, and between $L$ and symmetry energy, $E_{\rm sym}$ led to the limits at saturation density, $E_{\rm sym}=38.1\pm4.7$ MeV and $L(\rho_0)=106\pm37$ MeV~\citep{Reed2021}. \added{In addition, the recent S$\pi$RIT experiment at RIKEN suggested the slope of the symmetry energy to be $42\leq L \leq117$ MeV by using ratios of the charged pion spectra measured at high transverse momenta~\citep{RIKEN2021}.} This rather large value of $L$ from experimental data, as well as the constraint on $L$ from astrophysics, greatly challenge our understanding of nuclear many-body methods.
	
	Many attempts have been made to obtain the EoS of super-dense nuclear matter in NS under different nuclear many-body theoretical frameworks. The density functional theory (DFT) ~\citep{Vautherin1972,Shen1998,Douchin2001,Shen2002,Long2006,Long2007,Sun2008,Dutra2012,Bao2014a,Bao2014b} can effectively determine the nucleon-nucleon interaction by fitting the ground state properties of finite nuclei or the empirical saturation properties of infinite nuclear matter. It has been widely used to investigate various properties of finite nuclei and infinite nuclear matter and has achieved many successes in the fields of nuclear physics and astrophysics. Walecka proposed the first available version of covariant density functional theory (CDFT) based on the Hartree approximation~\citep{Walecka1974}, namely the $\sigma-\omega$ model, also known as the relativistic mean-field (RMF) model. Then the additions of $\rho$ meson, nonlinear terms of $\sigma$ and $\omega$ mesons, and the coupling terms of the $\rho$ meson with $\sigma$ and $\omega$ mesons improved this model and better described the nuclear many-body systems~\citep{Boguta1977,Serot1979,Sugahara1994,Horowitz2001}. 
	
	With the density increasing, many non-nucleonic degrees of freedom, like $\Delta$-resonances~\citep{Glendenning1991, Drago2014a,Drago2014b,Zhu2016,Li2018}, meson condensation ~\citep{Barshay1973,Baym1973,Pandharipande1995,Glendenning1999,Li-A2006} and hyperons ~\citep{Glendenning1985,Schffner1996,Shen2002,Weissenborn2012,Katayama2015} may appear in the inner core of NS from $(2-3)\rho_0~(\rho_0\approx0.16~\rm fm^{-3})$. The presence of hyperons will soften the EoS of NS matter, leading to a reduction in the maximum mass, which may be incompatible with the existence of stars with masses larger than $2M_{\odot}$. In the phase diagram of dense matter~\citep{Fukushima2011}, the hadronic phase plays a denominated role at low temperature and low chemical potential, while at high temperature or large chemical potential, the dense matter will be deconfined as quark-gluon plasma (QGP) in which the fundamental degrees of freedom are quarks and gluons. It has been proposed that in dense matter the quarks will gradually emerge as the density increases, leading to a hadron-quark phase transition~\citep{Baym1979,Celik1980,Satz1998}. The possible astrophysical role of quarks in NSs has been discussed frequently since the proposal of the quark models~\citep{Baym1976,Buballa2005}. 
	
	In the description of the quark phase,  the phenomenological models of interacting quarks~\citep{Nambu1961a,Nambu1961b,Nakazato2008,Sagert2009} should be adopted to describe the inner core of NS since it is still very difficult to directly use the Quantum Chromodynamics (QCD) theory except at very high baryon chemical potentials, i.e., $\mu_B\geq(3-6)$ GeV, or high baryon densities, i.e., $\rho_B\geq(10-100)\rho_0$~\citep{Freedman1977,Freedman1978,Fraga2001}. In this paper, a three-flavor Nambu-Jona-Lasinio (NJL) model with vector repulsion~\citep{Hatsuda1994,Vogl1991,Rehberg1996,Buballa2005,Wu2017} is adopted,  which was firstly proposed by Nambu and Jona-Lasinio in 1961~\citep{Nambu1961a,Nambu1961b}. The NJL model can successfully describe the dynamical chiral symmetry breaking and the generation of constituent quark masses. It has been widely used to study the role of quark degree of freedom in compact stars ~\citep{Schertler1999,Blaschke2005,Lawley2006,Klahn2007,Masuda2013-ptep,Masuda2013-apj,Orsaria2014,Chu2015}. Meanwhile, the vector repulsive interaction between quarks is taken into account since it can temper the growth of vector density and smooth out the chiral restoration~\citep{Kitazawa2002,Bratovic2013}.
	
	Different constructions have been employed to describe the hadron-quark phase transition in a neutron star, such as the Maxwell~\citep{Rosenhauer1991,Klahn2007,Agrawal2010} or Gibbs construction~\citep{Schertler1999,Li-A2009,Orsaria2013,Wu2017,Wu2018,Ju2021a,Ju2021b} for describing the first-order transition and interpolation constructions for converting the transition to be a crossover type~\citep{Masuda2013-ptep,Masuda2013-apj,Kojo2015}, where the hadronic and quark descriptions do not overlap. 
	Once the behaviors of NS matter at low and high densities are known, the EoS of the intermediate region where the phase transition occurs can be parameterized using an interpolating method to produce a massive neutron star. Masuda $et~al$~\citep{Masuda2013-ptep,Masuda2013-apj} investigated
	two different interpolation constructions, pressure, and energy density interpolated as hyperbolic functions of baryon density, respectively. Hell and Weise~\citep{Hell2014} interpolated pressure as a function of energy density using a similar function and Kojo $et~al$~\citep{Kojo2015} interpolated the pressure as a polynomial function of baryon chemical potential. However, these interpolation constructions are highly dependent on interpolation functions, leading to great uncertainty in EoSs obtained by interpolation functions.
	
	Therefore, in this work, we try to find an interpolation method that can reduce the uncertainty of interpolated values and give the magnitude of the uncertainties. Gaussian process regression (GPR) method from the statistics provides a good scheme to predict the unknown data with the training database and can give the uncertainty of the predictions~\citep{Rasmussen1996,Kuss2005,Rasmussen2006,Schulz2018}. Furthermore, we would like to determine the values of $L$  and the crossover windows with the latest observables from gravitational wave detection, NICER, and neutron skin thickness from PREX-\uppercase\expandafter{\romannumeral2}. In the hadronic phase, $\omega$ meson and $\rho$ meson coupling terms are included in the RMF model, i.e., the IUFSU parameter set~\citep{Fattoyev2010a,Fattoyev2010b}.  The coupling constants relevant to the vector–isovector meson, $\rho$, can be manipulated to generate different $L$~\citep{Dutra2012} at saturation density with fixed symmetry energy at subsaturation density~\citep{Horowitz2001,Zhang2013,Bao2014a}.

	This paper is organized as follows. In Sec.~\ref{theorfram} we briefly describe the RMF formalism, the NJL model and the GPR method. In Sec.~\ref{results}, properties of nuclear matter, and neutron stars are presented and discussed. Finally, the summary and conclusion will be given in Sec.~\ref{summary}.

\section{EQUATION OF STATE}\label{theorfram}
\subsection{The Hadronic Phase}
	
We adopt the relativistic mean-field (RMF) model to describe the hadronic matter in a low-density region. In the RMF model~\citep{Walecka1974,Muller1996,Horowitz2001,Bao2014a}, the nucleons interact with each other by exchanging various mesons, including the scalar-isoscalar meson($\sigma$), vector-isovector meson($\omega$), and vector-isovector meson($\rho$). The interaction between the $\omega$ and $\rho$ mesons is involved in a full Lagrangian,

	\begin{equation}
	\begin{aligned}
	\mathcal{L}_{\rm RMF}=& \sum_{i=n,p}\bar{\psi}_i\left\{i\gamma^{\mu}\partial_{\mu}-\left(M_i-g_{\sigma}\sigma\right)-\gamma^{\mu}\left(g_{\omega}\omega_{\mu}+\frac{g_{\rho}}{2}\vec{\tau}\vec{\rho_\mu}\right)\right\}\psi_i\\
	&+\frac{1}{2}\partial^{\mu}\sigma\partial_{\mu}\sigma-\frac{1}{2}m_{\sigma}^2\sigma^2-\frac{1}{3}g_{2}\sigma^3
	-\frac{1}{4}g_{3}\sigma^4\\
	&-\frac{1}{4}W^{\mu\nu}W_{\mu\nu}+\frac{1}{2}m_{\omega}^2\omega^{\mu}\omega_{\mu}+\frac{1}{4}c_3\left(\omega^{\mu}\omega_{\mu}\right)^2\\
	&-\frac{1}{4}\vec{R}^{\mu\nu}\vec{R}_{\mu\nu}{+}\frac{1}{2}m_{\rho}^2\vec{\rho}^{\mu}\vec{\rho}_{\mu}
	+\Lambda_{\rm v}\left(g_{\omega}^2\omega^{\mu}\omega_{\mu}\right)\left(g_{\rho}^2\vec{\rho}^{\mu}\vec{\rho}_{\mu}\right),
	\end{aligned}
	\end{equation}

where $W_{\mu\nu}$ and $\vec{R}_{\mu\nu}$ are the antisymmetry tensor fields of $\omega$ and $\rho$ mesons. Within the mean-field approximation, the meson fields are treated as classical fields, $\left\langle\sigma\right\rangle=\sigma,~\left\langle\omega_{\mu}\right\rangle=\omega,~\left\langle\vec{\rho}_{\mu}\right\rangle=\rho.$ Together with the Euler-Lagrange equations, the equations of motion for nucleons and mesons are given by

	\begin{equation}
	\begin{aligned}
	\left[i\gamma^{\mu}\partial_{\mu}-M_i^{*}-\gamma^0\left(g_{\omega}\omega
	+\frac{g_{\rho}}{2}\tau_{3}\rho\right)\right]\psi_i&=0,\\
	m_{\sigma}^2\sigma+g_{2}\sigma^2+g_{3}\sigma^3&=g_{\sigma}\left(\rho_n^s+\rho_p^s\right),\\
	m_{\omega}^2\omega+c_{3}\omega^3+2\Lambda_{\rm v}g_{\omega}^2g_{\rho}^2\omega \rho^2&=g_{\omega}\left(\rho_n+\rho_p\right),\\
	m_{\rho}^2\rho+2\Lambda_{\rm v}g_{\omega}^2g_{\rho}^2\omega^2\rho&=\frac{g_{\rho}}{2}\left(\rho_n-\rho_p\right).
	\end{aligned}
	\end{equation}
where $\rho_i^s,~\rho_i$ ($i=n,~p$) are the scalar and vector densities of species $i$, respectively. They are generated by the expectation value of nucleon fields.  $M_p^{*}=M_n^{*}=M_i-g_{\sigma}\sigma$ is the effective nucleon mass. 
	
	The hadronic matter in a neutron star which containing nucleons and leptons should satisfy the charge neutrality, $\rho_p=\rho_e+\rho_{\mu}$, and $\beta$ equilibrium, $\mu_p=\mu_n-\mu_e,~\mu_{\mu}=\mu_e$. The chemical potentials of nucleons and leptons can be derived from the thermodynamics equations at zero temperature,
	
	\begin{equation}
	\begin{gathered}
	\mu_i=\sqrt{k_{Fi}^2+M_i^{*2}}+g_{\omega}\omega+\frac{g_{\rho}}{2}\tau_3^i\rho,~~~~~i=n,~p;\\
	\mu_l=\sqrt{k_{Fl}^2+m_l^{2}},~~~~~l=e,~\mu,
	\end{gathered}
	\end{equation} 
   where $k_{Fi}$ is the Fermi momentum, which is related to the vector density by $\rho_i=k_{Fi}^3/3\pi^2$. With the energy-momentum tensor in a uniform system, the total energy density and pressure of the hadronic matter can be written as
   
	\begin{align}
	\mathcal{E}_{\rm HP}=&\frac{1}{\pi^2}\sum_{i=n,p}\int_0^{k_{Fi}}\sqrt{k_{Fi}^2+M_i^{*2}}k^2{\rm d}k+\frac{1}{2}m_{\sigma}^2\sigma^2+\frac{1}{3}g_{2}\sigma^3+\frac{1}{4}g_{3}\sigma^4\nonumber\\
	&+\frac{1}{2}m_{\omega}^2\omega^2+\frac{3}{4}c_{3}\omega^4+\frac{1}{2}m_{\rho}^2\rho^2+3\Lambda_{\rm v}\left(g_{\omega}^2\omega^{2}\right)\left(g_{\rho}^2\rho^{2}\right).\\ 
	P_{\rm HP}=&\frac{1}{3\pi^2}\sum_{i=n,p}\int_0^{k_{Fi}}\frac{k^4{\rm d}k}{\sqrt{k_{Fi}^2+M_i^{*2}}} 
	-\frac{1}{2}m_{\sigma}^2\sigma^2-\frac{1}{3}g_{2}\sigma^3-\frac{1}{4}g_{3}\sigma^4\nonumber\\
	&+\frac{1}{2}m_{\omega}^2\omega^2+\frac{1}{4}c_{3}\omega^4+\frac{1}{2}m_{\rho}^2\rho^2+\Lambda_{\rm v}\left(g_{\omega}^2\omega^{2}\right)\left(g_{\rho}^2\rho^{2}\right).
	\end{align} 
	
	\subsection{The Quark Phase}
	 In the description of quark matter, a three-flavor NJL model is adopted~\citep{Hatsuda1994,Masuda2013-apj}. The Lagrangian density is
	 
	\begin{equation}
	\begin{aligned}\label{1.3f.Lagrangian}
	\mathcal{L}_{\rm NJL}=&\bar{q}\left(i\gam^{\mu}\partial_{\mu}-m 
	\right)q+G_S\sum_{a=0}^8\left[(\bar{q}\lambda_a 
	q)^2+(\bar{q}i\gam_5\lambda_a  q)^2\right]{-G_V\left(\bar{q}\gam^{\mu}q\right)^2}\\
	&-K\left\{{\rm det}[\bar{q}(1+\gam_5)q)+{\rm 
		det}(\bar{q}(1-\gam_5)q]\right\}.
	\end{aligned}
	\end{equation}
	where $q$ is the quark field with three flavors and three colors together with the current quark mass matrix $m={\rm diag}(m_u,~m_d,~m_s)$. The term related to $G_S$ is a chiral symmetric four-quark interaction, where $\lambda_a$ are the Gell-Mann matrices with $\lambda_0=\sqrt{2/3}\bm{I}$. The term proportional to $G_V$ introduces additional vector and axial-vector interactions to produce universal repulsion between quarks~\citep{Kitazawa2002,Bratovic2013,Masuda2013-ptep}, which plays an important role in describing massive stars. The last term related to the coefficient $K$ corresponds to the six-quark Kobayashi-Maskawa-'t Hooft(KMT) interaction.
	
	Within mean-field approximation, the non-diagonal components of the condensates in a flavor space can be ignored. The constituent quark masses $m_i^{*}~(i=u,~d,~s)$ can be generated self-consistently through the gap equations,
	
	\begin{equation}
	m_i^{*}=m_i-4G_S\sigma_i+2K\sigma_j\sigma_k,
	\end{equation}
	where $\sigma_i=\langle\bar{q}_iq_i\rangle$ is the quark condensation in  $i-$flavor. The pressure can be evaluated from the thermodynamics potential, $P=-\Omega=\sum_i\mu_in_i-T\mc{S}-\varepsilon$, where $\varepsilon$ is the energy density, $T$ is the temperature, $\mc{S}$ is the entropy density, and $\mu_i$ is the effective chemical potential of quarks, which can be expressed as
	
	\begin{gather}
	\mu_i=\mu_i^{*}{+2G_V\sum_{i=u,d,s}n_i}.
	\end{gather}
	Here $n_i=\left\langle q_i^{\dag}q_i\right\rangle$ is the quark number density. The pressure and the energy density of quark phase from NJL model are
	
	\begin{align}
	P_{\rm NJL}=&-2G_S\left(\sig_u^2+\sig_d^2+\sig_s^2\right)+{G_V\left(n_u+n_d+n_s\right)^2}+4K\sig_u\sig_d\sig_s\nonumber\\
	&+\frac{3}{\pi^2}\sum_{i=u,d,s}\int_{k_{Fi}}^{\Lambda}k^2{\rm d}k\sqrt{k^2+m_i^{*2}}
	+\sum_{i=u,d,s}\mu_i^*\frac{k_{Fi}^3}{\pi^2}+\Omega_{\rm vac},\\
	\varepsilon_{\rm }=&2G_S\left(\sig_u^2+\sig_d^2+\sig_s^2\right){-G_V\left(n_u+n_d+n_s\right)^2}-4K\sig_u\sig_d\sig_s\nonumber\\
	&-\frac{3}{\pi^2}\sum_{i=u,d,s}\int_{k_{Fi}}^{\Lambda}k^2{\rm d}k\sqrt{k^2+m_i^{*2}}
	+\sum_{i=u,d,s}\frac{k_{Fi}^3}{\pi^2}(\mu_i-\mu_i^*)-\Omega_{\rm vac}.
	\end{align}
	where $\Omega_{\rm vac}$ is introduced to ensure that the pressure and energy density in vacuum are zero,
	
	\begin{gather}
	\Omega_{\rm 
		vac}=\sum_{i=u,d,s}\left[-\frac{3}{\pi^2}\int_0^{\Lambda}k^2{\rm d}k 
	\sqrt{k^2+{m_i^{*2}}}\right]+2G_S\left(\sig_u^2+\sig_d^2+\sig_s^2\right)-4K\sig_u\sig_d\sig_s.
	\end{gather}
	
	For the quark matter consisting of a neutral mixture of quarks ($u,~d,~s$) and leptons ($e,~\mu$), the $\beta$ equilibrium and the charge neutrality conditions should also be considered, 
	
	\begin{gather}
	\mu_s=\mu_d=\mu_u+\mu_e,~~~\mu_{\mu}=\mu_e.\\
	\frac{2}{3}n_u-\frac{1}{3}(n_d+n_s)-n_e-n_{\mu}=0.
	\end{gather}
	The total energy density and pressure of quark phase including the contributions from both  quarks and leptons are given by
	
	\begin{gather}
	\varepsilon_{\rm QP}=\varepsilon_{\rm 
		NJL}+\sum_{l=e,\mu}\int_{0}^{k_{Fl}}k^2{\rm d}k\sqrt{k^2+m_l^2},\\
	P_{\rm QP}=\sum_{i=u,d,s,e,\mu}n_i\mu_i-\varepsilon_{\rm QP}.
	\end{gather}

	\subsection{Hadron-quark Crossover}
	To describe the hadron-quark crossover in the NS, we need to construct an interpolation for $P(\rho_{B})$  so that the EoSs between hadron and quark matters can be smoothly connected. The crossover region is characterized by the lower density, $\rho_{L}$ and upper density, $\rho_{U}$. The EoSs of hadronic and quark matter are only used at $\rho_B<\rho_{L}$ and $\rho_B>\rho_{U}$ regions, respectively. In the density range $\rho_{L}<\rho_B<\rho_{U}$, the Gaussian Process Regression (GPR) method is adopted to perform the $P-\rho_B$ interpolation instead of using the pure hadronic matter EoS or the quark matter EoS. 

	Before introducing the Gaussian process regression method, we present a conventional univariate (one-dimensional) continuous probability distribution, Gaussian distribution, whose probability density function at random real variable $x$ in a set $X$ is
	
	\begin{gather}
	p(X=x)=\frac{1}{\sqrt{2\pi\sigma^2}}{\rm exp}\left[-\frac{\left(x-\mu\right)^2}{2\sigma^2}\right].
	\end{gather}
	The parameter $\mu$ is the mean value of the distribution, while $\sigma$ is the variance. $\sqrt{2\pi\sigma^2}$  is the normalization constant to ensure the integration of this distribution to be one. The expression, $X\sim\mc{N}(\mu,~\sigma^2)$ denotes that $p(X=x)=\mc{N}(x|\mu,~\sigma^2)$ and we can say that the variable in the set $X$ follows a Gaussian distribution.
	 However, a system is usually described by more than one feature variable, $(x_1,~\cdots,~x_n)$, that are correlated to each other. A multivariate Gaussian distribution should be used to model the variables altogether. The probability density function with an $n$-dimensional random vector $\bm{x}=(x_1,~\cdots,~x_n)^T$ can be extended as,
	 
	\begin{equation}\label{multiGP}
	\mc{N}(\bm{x}|\bm{\mu}, \bm{\Sigma})=(2\pi)^{-\frac{n}{2}}|\bm{\Sigma}|^{-\frac{1}{2}}{\rm exp}\left[-\frac{1}{2}(\bm{x-\mu})^T\bm{\Sigma}^{-1}(\bm{x-\mu})\right],
	\end{equation} 
	where $\bm{\mu}=(\mu_1,~\cdots,~\mu_n)^T$ represents the mean vector of $\bm{x}$ and $\bm{\Sigma}$ is an $n\times n$ {covariance} matrix. The $\bm{\Sigma}$ is defined to be a symmetric, positive-definite matrix with $\Sigma_{ij} = {\rm cov}(x_i,~x_j)$ for the $(i, j)$ element. Eq.\eqref{multiGP} can be represented as $\bm{X}\sim\mc{N}(\bm{\mu},~\bm{\Sigma})$. 
	
	 A Gaussian Process (GP) defined by Rasmussen is a collection of random variables, any finite numbers of which have the consistent joint Gaussian distributions~\citep{Rasmussen1996,Rasmussen2006}. More specifically, the multivariate Gaussian distribution describes the behavior of a finite random vector, while a Gaussian process is a stochastic one defined over a continuum of values{, which can be treated as a function}. In other words, in this process, a function ${f(\bm{x})}$ can be treated like a very long vector. It is fully defined by a mean function $m(\bm{x})$ and covariance function, $k(\bm{x},~\bm{x'})$, which respectively describe the mean value in the process at any point, and the covariance at any two points,
	 
	\begin{equation}\label{GP}
	f(\bm{x})\sim \mc{GP}\left(m(\bm{x}),~k(\bm{x},~\bm{x'})\right).
	\end{equation} 
	 For any finite set of points, the regression function modeled by a multivariate Gaussian distribution is given as
    $p(\bm{f}|\bm{X})=\mc{N}(\bm{f}|\bm{\mu},~\bm{K})$, which can be denoted by 
    
	\begin{equation}\label{N}
	\bm{f}=\left(f(x_1),~f(x_2),~\cdots,~f(x_n)\right)^T\sim \mc{N}\left(\bm{\mu},~\bm{K}\right),
	\end{equation}
	where $\bm{K}$ is the covariance matrix with elements $K_{ij} = k(x_i,~x_j)$ and $\bm{\mu}=\left(m(x_1),~\cdots,~m(x_n)\right)^T$ is the mean value vector. To clearly distinguish the Gaussian process and Gaussian distribution, we use $m$ and $k$ in the former and $\bm{\mu}$ and $\bm{K}$ in the latter. 
	 
	The covariance function, which is also called the kernel of the Gaussian process, can represent some form of distance or similarity~\citep{Murphy2012}. If the inputs $x_i$ and $x_j$ are close to each other, we generally expect that $f(x_i)$ and $f(x_j)$ will be close as well. This measure of similarity is embedded in the covariance function. There have been significant works on constructing kernels and analyzing their properties. The popular squared exponential (SE) covariance function is widely used and has the form
	
	\begin{equation}\label{SE}
	k_{\rm SE}(\bm{x},~\bm{x'})=s_f^2{\rm exp}\left(-\frac{||\bm{x}-\bm{x'}||^2}{2l^2}\right).
	\end{equation}
	The length-scale hyperparameter $l$ determines how wiggly the functions are. The signal variance parameter $s_f^2$ can be regarded as the amplitude, which controls the variability of sample functions from the mean function. The other available options for kernels, like the Rational Quadratic kernel, Mat$\rm\acute{e}$rn kernel, and $\gamma-$ exponential kernel are shown in Ref.~\citep{Rasmussen2006} and Ref.~\citep{Murphy2012}.
	
	Although a function can be treated as an infinite vector, we only have to make the predictions for finite points. Suppose a training set $\bm{X}=(x_1,~\cdots,~x_m)^T$ and the function outputs $\bm{y}=\left(f(x_1),~\cdots,~f(x_m)\right)^T$ with $\bm{y}\sim\mc{N}(\bm{\mu}_y,~\bm{K}_{yy})$.  Given a test set $\bm{X^*}=(x_1^*,~\cdots,~x_m^*)^T$, the regression function outputs  can be predicted by a multivariate Gaussian, $\bm{f}\sim\mc{N}(\bm{\mu}_f,~\bm{K}_{ff})$. According to the definition of the GP, previous observations $\bm{y}$ and the function values $\bm{f}$ follow a joint or multivariate normal distribution, which can be written as
	
	\begin{equation}\label{joint}
	\left(\begin{array}{c}
	\bm{y}\\
	\bm{f}\\
	\end{array}\right)
	\sim 
	\mc{N}\left(
	\left[\begin{array}{c}
	\bm{\mu}_y\\
	\bm{\mu}_f\\
	\end{array}\right]
	,~
	\left[\begin{array}{ccc}
	\bm{K}_{yy}  & \bm{K}_{yf}\\
	\bm{K}_{fy}  & \bm{K}_{ff}\\
	\end{array}\right]
	\right),
	\end{equation}            
	where $\bm{K}_{yy}=k(\bm{X},~\bm{X})$ is $n\times n$, $\bm{K}_{yf}=k(\bm{X},~\bm{X^*})$ is $n\times m$,  $\bm{K}_{fy}=\bm{K}_{yf}^T=k(\bm{X^*},~\bm{X})$ is $m\times n$, and $\bm{K}_{ff}=k(\bm{X^*},~\bm{X^*})$ is $m\times m$. The joint probability distribution over $\bm{y}$ and $\bm{f}$ is $p(\bm{y},~\bm{f}|\bm{X},~\bm{X^*})$, while the 
	regressions only need the conditional distribution $p(\bm{f}|\bm{y},\bm{X},~\bm{X^*})=\mc{N}(\bm{f}|\bm{\mu}_{f|y},~\bm{\Sigma}_{f|y})$ over the unknown outputs $\bm{f}$. Using the standard results, then we can obtain,
	
	\begin{equation}
	 \label{mu_Sigma}
	 \begin{aligned}
		\bm{\mu}_{f|y}&=\bm{\mu}_f+\bm{K}_{yf}^T\bm{K}_{yy}^{-1}(\bm{y}-\bm{\mu}_y),\\
	    \bm{\Sigma}_{f|y}&=\bm{K}_{ff}-\bm{K}_{yf}^T\bm{K}_{yy}^{-1}\bm{K}_{yf}.
 	 \end{aligned}
	\end{equation}
	Furthermore, a noise is often taken into account in the GPR models. If there is an additive independent Gaussian noise, $\epsilon$ with variance $\sigma_{\epsilon}^2$, the output $\bm{y'}$ of a function $f$ at input $\bm{X}$ can be written as 
	
	\begin{equation}
	\bm{y'}=f(\bm{x})+\epsilon,~~~\epsilon\sim\mc{N}\left(0,~\sigma_{\epsilon}^2\right).
	\end{equation} 
	The prior on the noisy observations becomes ${\rm cov}(\bm{y'})=\bm{K}_{yy}+\sigma^2\bm{I}$. Then the joint distribution of the noisy observations and the function values at testing point becomes
	
	\begin{equation}\label{joint_noisy}
	\left(\begin{array}{c}
	\bm{y'}\\
	\bm{f}\\
	\end{array}\right)
	\sim 
	\mc{N}\left(0,~
	\left[\begin{array}{ccc}
	\bm{K}_{yy}+\sigma_{\epsilon}^2\bm{I}  & \bm{K}_{yf}\\
	\bm{K}_{fy}  & \bm{K}_{ff}\\
	\end{array}\right]
	\right),
	\end{equation}
	where the mean function is zero for notational simplicity.  Hence the posterior predictive equation in GPR is
	
	\begin{gather} \label{cond}
	\bm{f}|\bm{y'},\bm{X},~\bm{X^*}\sim\mc{N}\left(\bm{K}_{yf}^T\left(\bm{K}_{yy}+\sigma_{\epsilon}^2\bm{I}\right)^{-1}\bm{y'},~
	\bm{K}_{ff}-\bm{K}_{yf}^T\left(\bm{K}_{yy}+\sigma_{\epsilon}^2\bm{I}\right)^{-1}\bm{K}_{yf}~\right).
	\end{gather}
	 
The chosen kernel can decide the predictive performance of GP. In this work, we choose the SE kernel in Eq.\eqref{SE} for the noisy observations. The kernel function contains hyperparameters such as the length-scale, signal variance, and noise variance, which need to be inferred from the known data. The inferences for all hyperparameters should be obtained by computing the probability of the known data instead of figuring out the contributions of hyperparameters. A common approach is to estimate them by taking the maximum marginal likelihood. Given the hyperparameters $\theta=l,~s_f,~\sigma_{\epsilon}$ and inputs $(\bm{X},~\bm{y'})$, from Eq.\eqref{cond}, the negative log marginal likelihood can be written as  
 \begin{equation} 
      \log p(\bm{y'}|\theta,~\bm{X})=-\frac{1}{2}\bm{y'}^T\bm{K}_{y'}^{-1}\bm{y'}
      -\frac{1}{2}\log\big|\bm{K}_{y'}\big|-\frac{n}{2}\log(2\pi),~~~\theta=l,~s_f,~\sigma_{\epsilon},
\end{equation}
where $\bm{K}_{y'}=\bm{K}_{yy}+\sigma_{\epsilon}^2\bm{I}$ is the covariance matrix for the noisy targets $\bm{y'}$. The first term in above equation measures the data fit, the second term is a model complexity term, and the last term is a normalization constant.  Now the  partial derivatives of negative log marginal likelihood with respect to the hyperparameters can be obtained 

	\begin{equation} \label{marg}
	\frac{\partial }{\partial\theta}\log p(\bm{y'}|\theta,~\bm{X})
	=\frac{1}{2}{\bm{y'}^T\bm{K}_{y'}^{-1}\frac{\partial \bm{K}_{y'}}{\partial\theta}\bm{K}_{y'}^{-1}\bm{y'}}
	-\frac{1}{2}{\rm tr}\left(\bm{K}_{y'}^{-1}\frac{\partial\bm{K}_{y'}}{\partial\theta}\right),~~~\theta=l,~s_f,~\sigma_{\epsilon}=0.
	\end{equation}
Finally, the outputs, $\bm{f}$ will be well defined after solving this equation to get the hyperparameters $l,~s_f,~\sigma_{\epsilon}$. 
	
When interpolating the pressure $P$ as a function of baryon density, $\rho_B$, in the hadron-quark crossover region of NS, we can treat the set $(\bm{\rho_B},~\bm{P})$ in hadronic and quark phases as observations, where $\bm{P}=f(\bm{\rho_B})$. The prior distribution on the noisy observations of a zero-mean Gaussian process is
$\bm{P}\sim\mc{N}(0,~k(\bm{\rho_B},~\bm{\rho_B})+\sigma_{\epsilon}^2\bm{I})$ with SE kernel. The  target point to be predicted is the pressure $\bm{P^*}$ at the corresponding $\bm{\rho_B^*}$ in the crossover density region. Then the joint distribution of training observations $\bm{P}$ and predictive targets $\bm{P^*}$ will be obtained from Eq.\eqref{joint_noisy}. Finally, we can use the {conditioning} and marginalization as Eq.\eqref{cond}-Eq.\eqref{marg} to find the best hyperparameters and figure out the mean values and the covariance, where the mean values are the pressures in the hadron-quark crossover section that we expect. \replaced{In addition, we also can calculate the $95\%$ confidence interval from the covariance to generate the uncertainties of the interpolation method.}{In addition, the $95\%$ confidence interval can be estimated by the standard deviation, which can be calculated through the diagonal element of the covariance as shown in Eq.\eqref{mu_Sigma} or Eq.\eqref{cond} to generate the uncertainties of the interpolation method.}

	\section{Results and Discussion}\label{results}
For the hadronic phase of NS, the RMF model with the IUFSU parameter set is employed. Its original parameter set and corresponding saturation properties of symmetric nuclear matter are given in Table.\eqref{table.IUFSU_set} and Table.\eqref{table.IUFSU_sat}, respectively~\citep{Fattoyev2010b}. There is an additional coupling term between $\omega$ and $\rho$ mesons compared to other RMF parameter sets, $\Lambda_{\rm v}$, which will influence the density dependence of symmetry energy. The slope of symmetry energy from the IUFSU model is just $47.165$ MeV. It is much smaller than those from the NL3 or TM1 parameter set, which is around $110$ MeV.  
	\begin{table}[htb]
		\footnotesize
		\centering
		\caption{IUFSU parameter set in the RMF model ~\citep{Fattoyev2010b}.}\label{table.IUFSU_set}
		\footnotesize
		\begin{tabular}{lccccccccccccccccc}
			\hline\hline
			&   &$M[\rm MeV]$  &$m_{\sigma}[\rm MeV]$ &$m_{\omega}[\rm MeV]$ &$m_{\rho}[\rm MeV]$ &$g_{\sigma}$ &$g_{\omega}$ &$g_{\rho}$  &$g_2[\rm fm^{-1}]$ &$g_3$ &$c_3$ &$\Lambda_{\rm v}$ \\ 
			\hline
			&IUFSU &939.0000 &491.5000 &782.5000 &763.0000 &9.9713 &13.0312  &13.5900 &8.4929 &0.4877 &144.2195 &0.046\\ 
			\hline\hline
		\end{tabular}
	\end{table}	
	
            \begin{table}[htb]
            	\small
            	\centering
            	\caption{Saturation properties of nuclear matter for the IUFSU model. $E_0,~K,~E_{\rm sym},~L$ are the energy per nucleon, incompressibility, symmetry energy, and the slope of symmetry energy at saturation density
            		$\rho_0$, respectively.}\label{table.IUFSU_sat}
            	\footnotesize
            	\begin{tabular}{lcccccccccccccc}
            		\hline\hline	
            		& &$\rho_0[\rm fm^{-3}]$ &$E_0[\rm MeV]$ &$K[\rm MeV]$  &$E_{\rm sym}[\rm MeV]$  &$L[\rm MeV]$ \\
            		\hline
            		&IUFSU &0.155 &-16.397 &230.749 &31.336  &47.165    \\ 
            		\hline\hline
            	\end{tabular}
            \end{table}
            
 Around the saturation density of symmetry nuclear matter, $\rho_0$, the symmetry energy, $E_{\rm sym}$, can be expanded in a Taylor series as a function of baryon vector density, $\rho_B=\rho_n+\rho_p$,
 
	\begin{equation}
	E_{\rm sym}(\rho_B)= E_{\rm sym}(\rho_0)+\frac{L}{3}\left(\frac{\rho_B-\rho_0}{\rho_0}\right)
	+\frac{K_{\rm sym}}{18}\left(\frac{\rho_B-\rho_0}{\rho_0}\right)^2+\cdots,
	\end{equation}
where the slope of the symmetry energy is $L=3\rho_0\left[\partial E_{\rm sym}(\rho_B)/\partial\rho_B\right]_{\rho_B=\rho_0}$. It characterizes the density dependence of $E_{\rm sym}$ and is linearly  correlated with the neutron-skin thickness $R_{\rm skin}^{208}$ of $^{208}{\rm Pb}$. However, the uncertainty in the present measurement, such as PREX experiments,~\citep{Abrahamyan2012,Adhikari2021}, prevents us from inferring the slope of the symmetry energy. In order to explore the influence of  $L$ on neutron-rich systems, several new parameter sets based on the original IUFSU set are developed, which can keep the isoscalar properties of nuclear matter and finite nuclei from the IUFSU set, and change their isovector properties by controlling the strengths of $g_\rho$ and $\Lambda_{\rm v}$~\citep{Bao2014a}. 
	
The parameters, $g_\rho$ and $\Lambda_{\rm v}$, based on the IUFSU model with $L$ from \replaced{$70$ to $110$ MeV}{$50$ MeV to $110$ MeV} are listed in Table\eqref{table.IUFSU_slope}~\citep{Bao2014a,Hu2020}. The parameters $g_\rho$ and $\Lambda_{\rm v}$ are adjusted, where $L$ is a given value at saturation density and $E_{\rm sym}$ is fixed at the nuclear sub-saturation density of $\rho=0.11~{\rm fm}^{-3}$. \replaced{These $L$ satisfy the recent constraints $L=106\pm 37$ MeV, which is obtained by using the strong correlations between $R_{\rm skin}^{208}$ and $L$ within the covariant energy density functional by PREX-II.}{These $L$ satisfy the recent constraints, $42\leq L \leq117$ MeV from S$\pi$RIT Collaboration~\citep{RIKEN2021} and $L=106\pm 37$ MeV obtained by using the strong correlations between $R_{\rm skin}^{208}$ and $L$ within the covariant energy density functional from PREX-II~\citep{Reed2021}.} Other parameters are completely same as those in the original IUFSU set given in Table ~\eqref{table.IUFSU_set}.  The symmetry energies at the saturation point and the corresponding $R_{\rm skin}^{208}$ from these parameter sets also satisfy the constraints from PREX-II, $E_{\rm sym}(\rho_0)=38.1\pm4.7~\rm MeV$~\citep{Reed2021}, and $R_{\rm skin}^{208}=0.283\pm 0.071~\rm fm$~\citep{Adhikari2021}.

   \begin{table}[htb]
	\footnotesize
	\centering
	\caption{ The parameters $g_{\rho}$ and $\Lambda_{\rm v}$ generated from the IUFSU model for different slope $L$ at saturation density $\rho_0$ with fixed symmetry $E_{\rm sym}=26.78~\rm MeV$ at $\rho=0.11~\rm fm^{-3}$. $E_{\rm sym}(\rho_0)$ and $R_{\rm skin}^{208}$ are the symmetry energy at saturation density, and the neutron skin thickness of $^{208}{\rm Pb}$, respectively~\citep{Bao2014a}.}\label{table.IUFSU_slope}
	\begin{tabular}{lccccccc}
		\hline\hline
		&$L$(MeV)                     ~~~&\added{50.0}     ~~~&70.0   ~~~&90.0   ~~~&110.0\\
		\hline
		&$g_{\rho}$                   ~~~&\added{12.8202}  ~~~&10.3150 ~~~&9.3559 ~~~&8.8192 \\
		&$\Lambda_{\rm v}$            ~~~&\added{0.0420}  ~~~&0.0220  ~~~&0.0098 ~~~&0.0011 \\
		&$E_{\rm sym}(\rho_0)$[MeV]   ~~~&\added{31.68}     ~~~&33.94   ~~~&35.74  ~~~&37.27\\
		&$R_{\rm skin}^{208}$ [fm]    ~~~&\added{0.1739}   ~~~&0.2278  ~~~&0.2571 ~~~&0.2770\\
		\hline\hline
	\end{tabular}
\end{table} 

Considering an NS consisting of the pure hadronic matter, its properties, such as the maximum masses, $M_{\rm max}$, the corresponding radius, $R_{\rm max}$, the central density, $\rho_{\rm max}$, the radius of $1.4M_\odot$ NS, $R_{1.4}$, the density of $1.4M_\odot$ NS, $\rho_{1.4}$, and dimensionless tidal deformability of $1.4M_\odot$ NS, $\Lambda_{1.4}$ can be derived by solving the Tolman-Oppenheimer-Volkoff (TOV) equation~\citep{Tolman1939,Oppenheimer1939} with the EoS of neutron star matter. These properties of neutron stars obtained from the above IUFSU parameter sets with different $L$ are shown in Table.~\eqref{NS_IUFSU}. We can find that the maximum masses of NS from these sets are not sensitive to $L$ and are all less than $2M_\odot$. On the other hand, with the increase of $L$, the radii corresponding to the maximum mass are from \replaced{$11.24$ km to $11.71$ km}{$11.15$ km to $11.71$ km}, while the radii of $1.4M_\odot$ change in the range of \replaced{$12.74$ km to $13.54$ km}{$12.40$ km to $13.54$ km}. The central densities are around $\rm 1.0~fm^{-3}$. They become about $0.38-0.43\rm~fm^{-3}$ for $1.4M_{\odot}$ NS~\citep{Hu2020}.

\begin{table}[htb]
		\footnotesize
		\centering
		\caption{NS properties for IUFSU model with different $L$.}\label{NS_IUFSU}
		\begin{tabular}{lccccccccccc}
			\hline\hline
			$L$(MeV)              ~~~&\added{50.0}  ~~~&70.0 ~~~&90.0 ~~~&110.0  \\
			\hline
			$M_{\rm max}/M_{\odot}$       ~~~&\added{1.9387}   ~~~&1.9365   ~~~&1.9447   ~~~&1.9853  \\
			$R_{\rm max}[\rm km]$         ~~~&\added{11.1548}   ~~~&11.2419  ~~~&11.4126  ~~~&11.7106  \\
			$\rho_{\rm max}[\rm fm^{-3}]$ ~~~&\added{1.0251}    ~~~&1.0418   ~~~&1.0192   ~~~&0.9651   \\
			$R_{\rm 1.4}[\rm km]$         ~~~&\added{12.4008}  ~~~~&12.7370  ~~~&13.0800  ~~~&13.5400   \\
			$\rho_{\rm 1.4}[\rm fm^{-3}]$ ~~~&\added{0.4330}    ~~~&0.4285   ~~~&0.4158   ~~~&0.3862    \\
			$\Lambda_{\rm 1.4}$           ~~~&\added{512}       ~~~&532      ~~~&604      ~~~&735     \\
			\hline\hline
		\end{tabular}
	\end{table}
 
For the quark phase, the HK parameter set of SU(3) NJL model was adopted~\citep{Hatsuda1994} with $\Lambda=631.4$ MeV, $G_S\Lambda^2=1.835$, $G_D\Lambda^5=9.29$, $m_{u,d}=5.5$ MeV, and $m_s=135.7$ MeV, where $\Lambda$ is the three-momentum cutoff.  In SU(3) NJL model, the vector coupling $G_V$ has not been well determined. It has a similar magnitude to the scalar coupling scale $G_V\sim G_S$ or $G_V\sim 2G_S$~\citep{Bratovic2013,Masuda2013-apj} in order to explain the lattice results on the curvature of the linear chiral restoration at zero density. The studies on the QCD phase diagram suggest that it can be comparable to or even larger than $G_S$~\citep{Lourenco2012}. Therefore its value is chosen as $G_V=1.0,~1.4,~1.8G_S$ in this work to study its influences on quark matter since the vector repulsion plays an important role in stiffening the EoS. With the increase of $G_V$, the stiffer EoSs can be obtained and generate massive compact stars.

\begin{figure}[htb]
		\centering
		\includegraphics[scale=0.5]{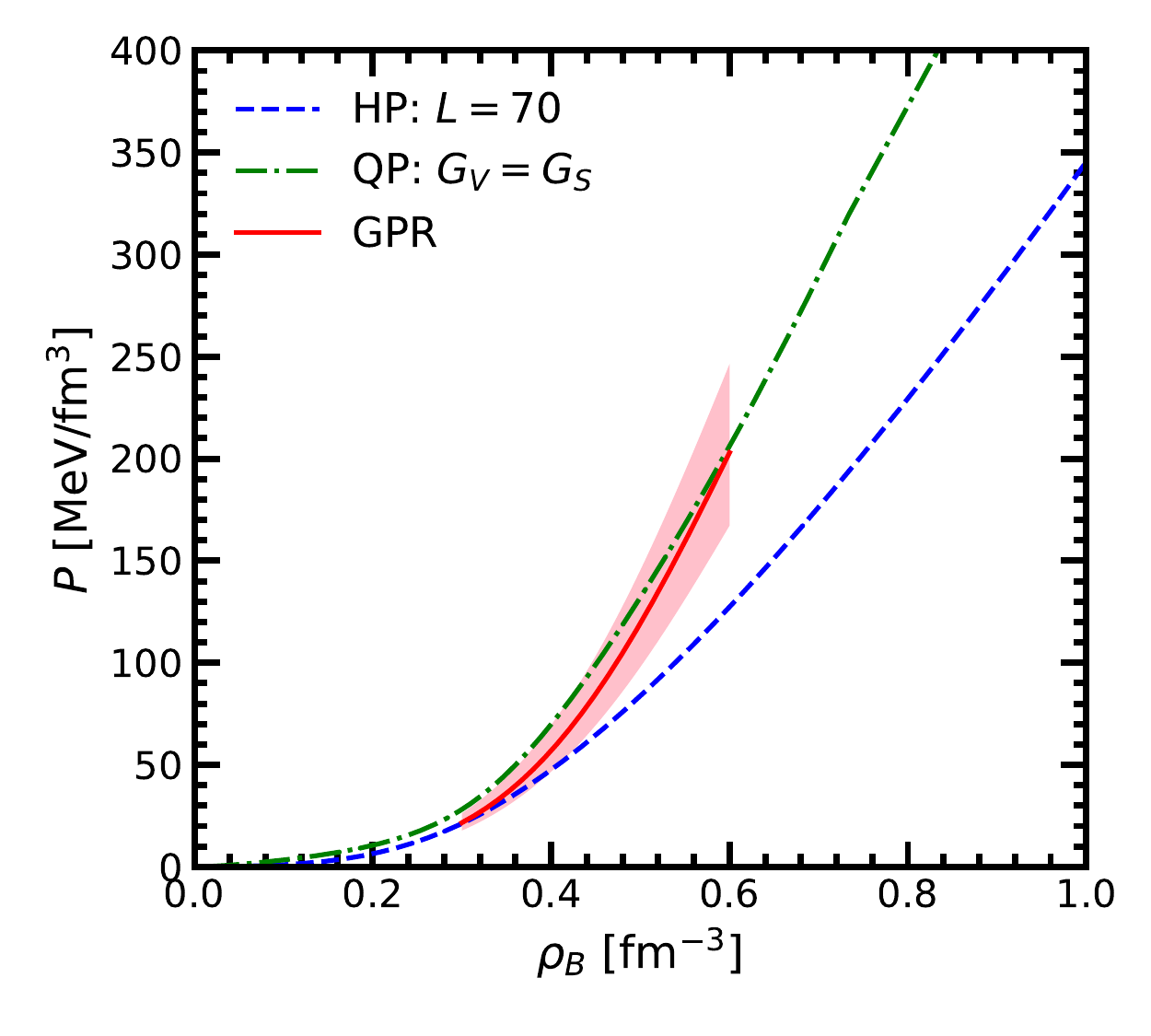}
		\caption{The interpolated pressure between IUFSU model with $L=70$ in hadronic phase and NJL model with $G_V=G_S$ in quark phase with GPR method.  \label{fig.GPR}}
	\end{figure}
	
\added{In the hadron-quark crossover region, the interpolation method is not unique. At least, there are three choices to connect the hadronic and quark phases for the crossover EoSs, which are performed in the $P-\rho_B$, $\varepsilon-\rho_B$, or $P-\mu_B$ plane~\citep{Masuda2013-ptep,Kojo2015,Baym2018}. In this work, we use the  $P-\rho_B$ combination, while  the quark phase is described by conventional NJL model including the vector coupling term, $G_V$. Actually, we previously attempted the other two combinations of $P-\mu_B$ and $\varepsilon-\rho_B$ before. However, the results were not reasonable for the EoSs of massive neutron stars with larger vector coupling strengths $G_V$. Therefore, the $P-\rho_B$ interpolation is considered in present work.}
An example of $P-\rho_B$ interpolation between hadronic phase (HP) and quark phase (QP) with the GPR method is shown in Fig.\eqref{fig.GPR}. The density range of the interpolation region, which is also called the crossover window, is from $0.3~\rm fm^{-3}$ to $0.6~\rm fm^{-3}$. The EoS of HP is provided from the IUFSU model with $L=70$ MeV, while that of QP is generated by the NJL model with $G_V=G_S$. The solid curve is the interpolation result with the GPR method, and the shaded area represents the $95\%$ confidence interval of the interpolated pressure. \replaced{The corresponding energy density,  $\varepsilon(\rho_B)$ can be calculated by integrating the thermodynamic relation $P={\rho_B}^2\partial(\varepsilon/\rho_B)/\partial{\rho_B}$ consistently}{The corresponding energy density,  $\varepsilon(\rho_B)$ can be obtained by integrating the thermodynamic relation $\partial\varepsilon/\textbf{}\rho_B=(P+\varepsilon)/\rho_B$ numerically with a given initial value, $\varepsilon_L$, following the idea in Ref.~\citep{Masuda2013-ptep,Masuda2013-apj}. This method guarantees the thermodynamic consistency in the present calculation.} 

\begin{figure}[htb]
	\centering
	\includegraphics[scale=0.3]{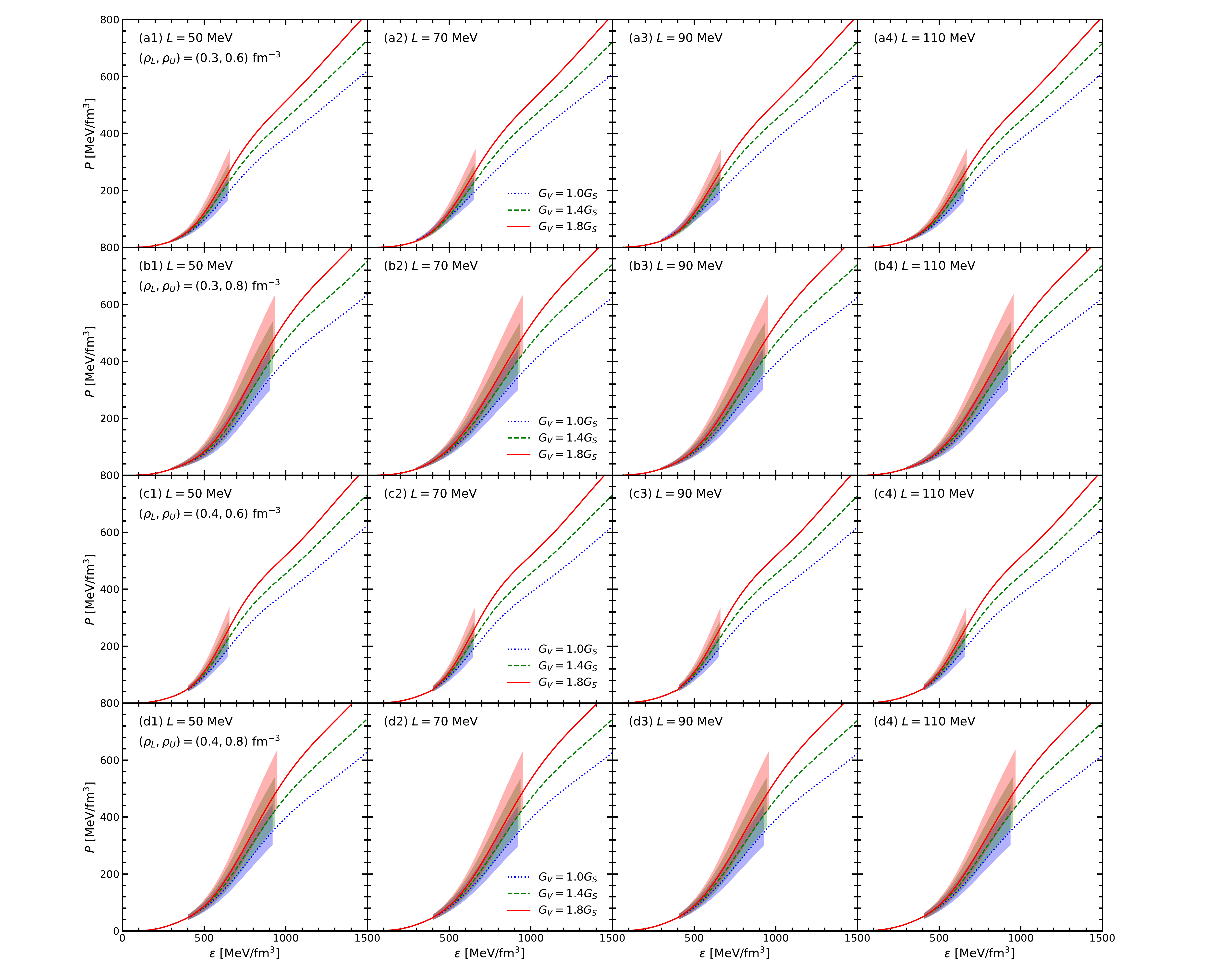}
	\caption{Interpolated pressure as a function of energy density with GPR interpolation
		between hadronic matter from family IUFSU parameter sets 
		and quark matter with $G_V=1.0,~1.4,~1.8G_S$. The crossover windows here are fixed to be 
		(a) $(\rho_L,~\rho_U)=(0.3,~0.6)\rm ~fm^{-3}$, 
		(b) $(\rho_L,~\rho_U)=(0.3,~0.8)\rm ~fm^{-3}$, 
		(c) $(\rho_L,~\rho_U)=(0.4,~0.6)\rm ~fm^{-3}$,
		(d) $(\rho_L,~\rho_U)=(0.4,~0.8)\rm ~fm^{-3}$.}\label{fig.GPR_P-e}
\end{figure}

When modeling phase transitions in NSs, the conventional choice for the apparent density of the hadron-quark crossover is $\rho_{L}=(2-3)\rho_0$, below which the hadronic phase exists. Similarly, the deconfinement density of nucleons should happen in the high-density region. The reasonable choice for the closure density of the hadron-quark crossover  is around $\rho_{U}\sim(4-7)\rho_0$~\citep{Baym1979,Celik1980}. To investigate the effects of the crossover window between hadronic and quark phases on the EoSs and NS structure, the crossover windows are fixed as $(\rho_{L},~\rho_{U})=(0.3,~0.6),~(0.3,~0.8),~(0.4,~0.6),~(0.4,~0.8)~\rm fm^{-3}$  in this work.  \added{In addition, the hyperonic star including the hyperons with different nonlinear and density-dependent models have been investigated in the framework of the relativistic mean-field model~\citep{Huang2022}, which include the IUFSU parameter set. The threshold density of the first hyperon for IUFSU model is about $0.38$ $\rm fm^{-3}$. The inclusion of hyperons will soften the EoS of the hadronic phase. When the hadron-quark crossover phase starts from $\rho_L=0.3$ fm$^{-3}$, the hyperon cannot appear in the hadronic phase, while the $\rho_L$ is changed to $0.4$ fm$^{-3}$, the hyperon only can exist between $0.38$ to $0.40$ fm$^{-3}$. It may slightly make the hadronic phase EoS softer. Therefore, the hyperon effect on the crossover EoS and the corresponding neutron structure can be negligible.}

In Fig.~\eqref{fig.GPR_P-e}, the pressure as a function of energy density are plotted in the case of $P-\rho_B$ interpolation with GPR method between IUFSU model with \replaced{$L=70,~90,~110$ MeV}{$L=50,~70,~90,~110$ MeV} and NJL model with $G_V=1.0,~1.4,~1.8G_S$. The crossover windows are chosen to be $(\rho_{L},~\rho_{U})=(0.3,~0.6),~(0.3,~0.8),~(0.4,~0.6),~(0.4,~0.8)~\rm fm^{-3}$ and can be inferred from the density range of the shaded areas. \replaced{These EoSs in the low-density region are mainly determined by different $L$ in the hadronic phase.}{The EoSs of neutron star crust as the nonuniform matter were calculated with the corresponding IUFSU parameterizations in the framework of the self-consistent Thomas-Fermi approximation~\citep{Bao2015}. In the core region of a neutron star, the EoSs of the uniform matter in the low-density region are mainly determined by different $L$ in the hadronic phase.} When hadron-quark crossover appears, the pressure and energy density are strongly dependent on the crossover windows and the vector coupling strength of quarks. A larger $G_V$ can generate a stiffer EoS. Furthermore, the uncertainties in the interpolation process can also be obtained. The shaded areas are $95\%$ confidence intervals. The uncertainty increases rapidly with pressure increasing and  it is about $20\%$ for $G_V=1.8 G_S$ to the pressure at the upper limit of the crossover window, $\rho_{U}$.

\begin{figure}[htb]
	\centering 
	\includegraphics[scale=0.3]{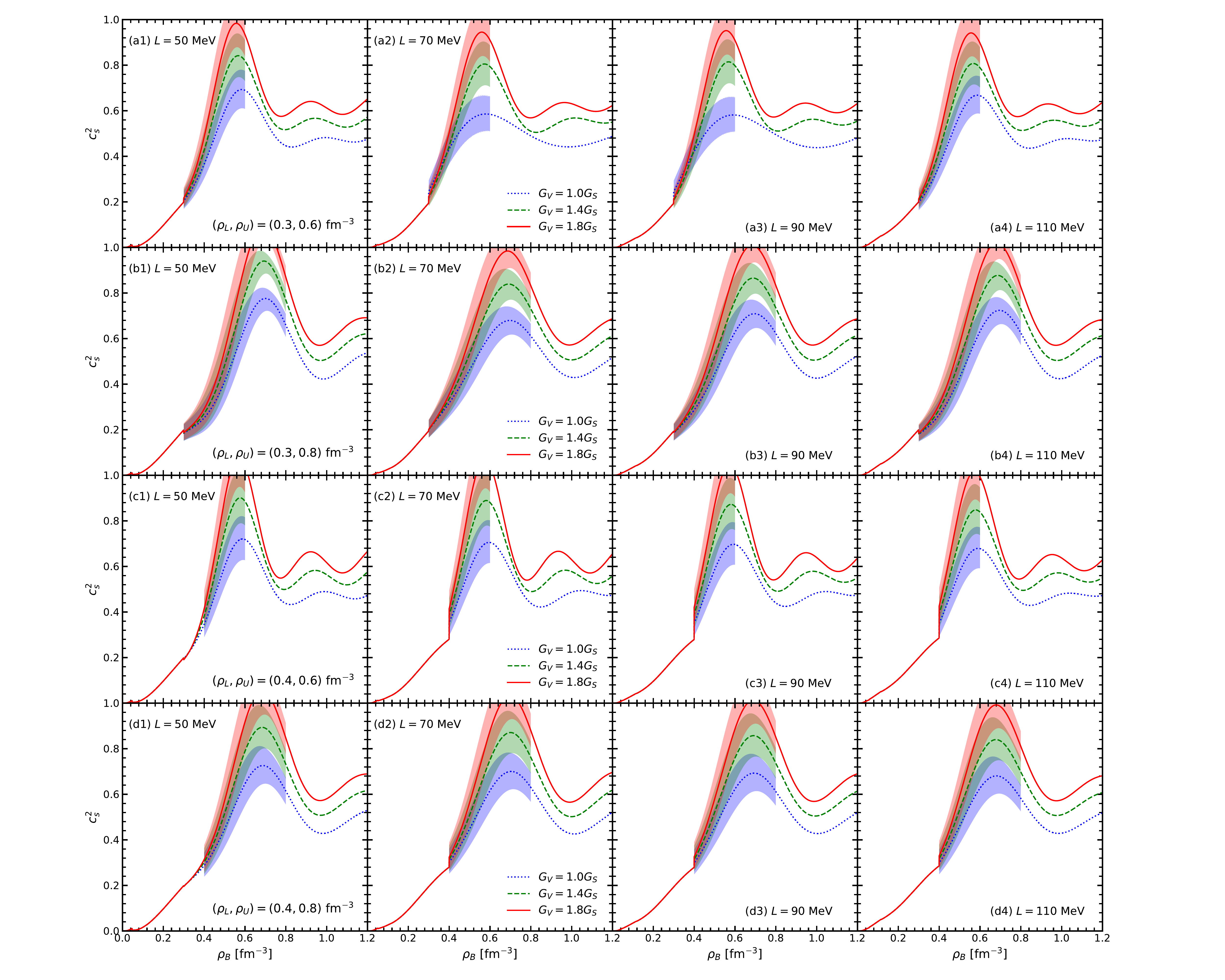}
	\caption{Speeds of sound vs. baryon density with GPR interpolation. The parameter sets in hadronic phase and quark phase, the crossover windows are the same as Fig.~\eqref{fig.GPR_P-e}.}\label{fig.GPR_v-nb}
\end{figure}			

The speed of sound, $c_s=\sqrt{{\rm d}P/{\rm d}\varepsilon}$ in NS matter is one of the measurements to quantify the stiffness of EoS. In Fig.~\eqref{fig.GPR_v-nb}, $c_s^2$ of the interpolated EoS with different $L$ and $G_V$ as a function of density are plotted.  A discontinuity  of $c_s^2$ at the lower limit density, $\rho_L$, is caused by the happening of hadron-quark crossover. It steeply increases at the beginning crossover region and approaches a maximum value around $1.0$ at $\rho=0.6~\rm fm^{-3}$ and then reduces to around $0.6$ at the high-density region.  Furthermore, it is found that the vector strength between quarks cannot be too strong, otherwise, the speed of sound of NS matter may exceed $1.0$, which will violate the causality. When the crossover window is fixed, the slope of symmetry energy has a few influences on the $c_s$. Its magnitude will be obviously changed with the crossover window at a fixed $L$. Since $c_s$ largely increases as $\rho_L$ increases, the hadron-quark crossover in NS cannot occur later than baryon density, $0.4~\rm fm^{-3}$ with a stiffer EoS, which is consistent with recent Bayesian analysis under the assumption of first-order hadron-quark phase transition and the join constraints from NICER and GW170817~\citep{Li2021}. Recently, we also explored the structured hadron-quark mixed phase with the energy minimization method, Gibbs construction, and Maxwell one, where the onset density of quark is strongly dependent on the quark interactions~\citep{Ju2021a, Ju2021b}.

\begin{figure}[htb]
	\centering 
	\includegraphics[scale=0.3]{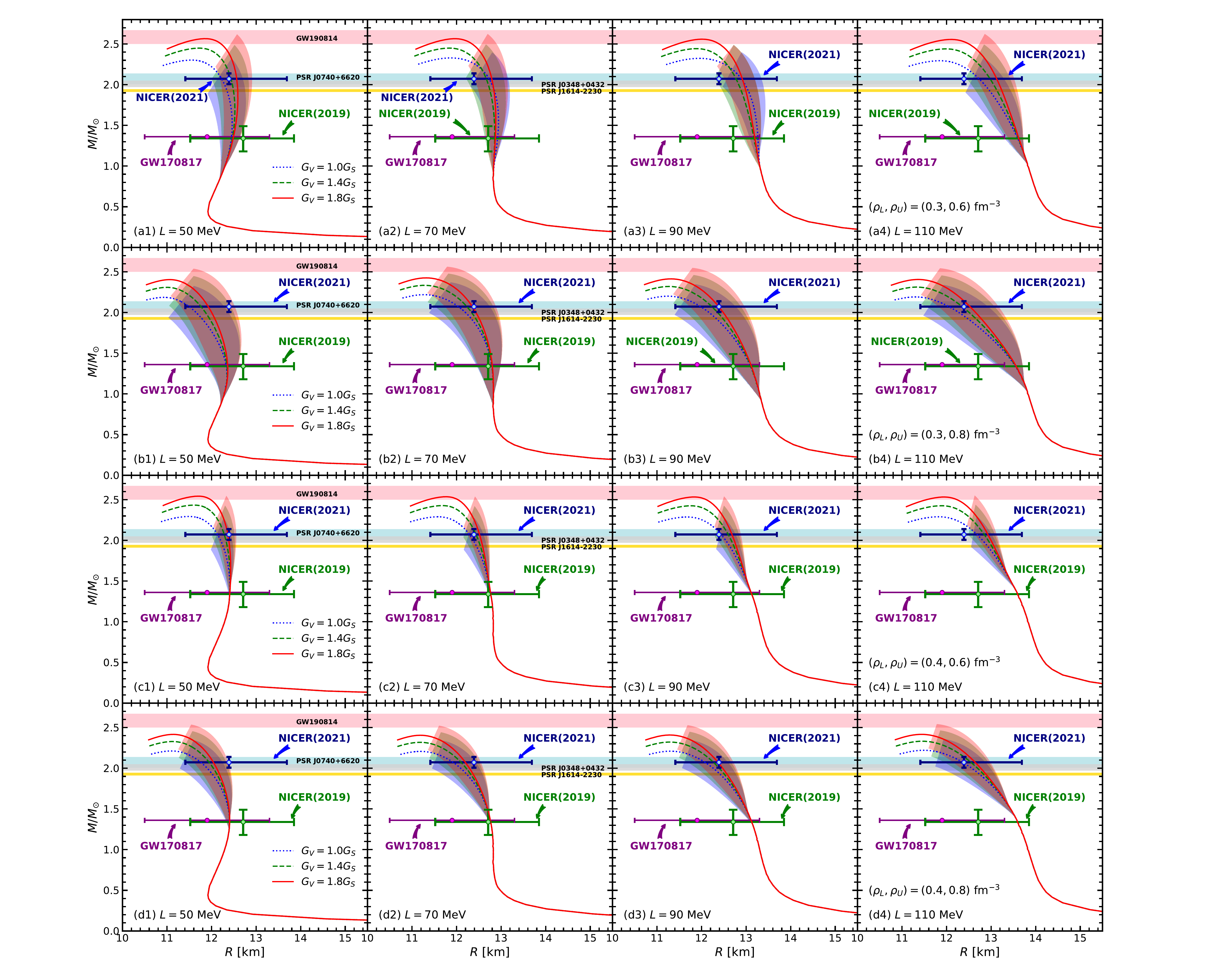}
	\caption{The Radius vs. Neutron star mass with the EoS from GPR interpolation method. The parameter sets in hadronic phase and quark phase, the crossover windows are the same as Fig.~\eqref{fig.GPR_P-e}.}\label{fig.GPR_rad-mass}
\end{figure}

The properties of neutron star, such as the mass-radius relation, can be provided  after substituting the EoS to TOV equation. In Fig.~\eqref{fig.GPR_rad-mass}, the mass-radius relations are shown within above EoSs to describe the hadron-quark crossover with GPR method. In addition, various constraints from the observables of massive neutron stars, PSR J1614-2230 ($1.928\pm 0.017M_{\odot}$)~\citep{demorest2010,antoniadis2013}, PSR J0348+0432 ($2.01\pm 0.04M_{\odot}$)~\citep{antoniadis2013}, the mass and radius simultaneous measurement of PSR J0740+6620 (a mass of  $2.072^{+0.067}_{-0.066}M_\odot$ with a radius $12.39_{-0.98}^{+1.30}$ km~\citep{Riley2021} ) and PSR J0030+0451 (a mass of $1.34_{-0.16}^{+0.15}M_{\odot}$ with a radius $12.71_{-1.19}^{+1.14}$ km~\citep{Riley2019} ) are shown. The radius of neutron star at $1.4 M_\odot$ extracted from GW170817, $R_{1.4}=11.9\pm1.4$ km~\citep{Abbott2018} is also considered. 

With the introduction of the hadron-quark crossover, the present EoSs can easily describe the $2M_\odot$ massive neutron star, while they are just around $1.95 M_\odot$ with the pure hadronic matter from IUFSU parameter sets as shown before. The maximum mass of NS is strongly dependent on $\rho_U$ of the crossover window and $G_V$ of quark interaction. It can be heavier than $2.55 M_\odot$ with $G_V=1.8G_S$ due to the stiffer EoS. Therefore, the secondary compact object in GW190814 may be an NS containing the hadron-quark crossover.  If the pure quark phase appears later than $0.6~\rm fm^{-3}$, the maximum mass of NS in the mean value will be less than $2.5 M_\odot$ since the hadronic matter can provide more contributions to the compositions of the NS. The radius of NS is very sensitive to the slope of symmetry energy $L$.  It will become larger with the $L$ increasing not only at $1.4 M_\odot$ but also at $2.0 M_\odot$. The cases of $L=110$ MeV are almost excluded by the constraint of gravitational wave detection, GW170817 event. The hadron-quark crossover can appear in the lower-mass NS around $1.0M_\odot$ with $\rho_{L}=0.3~\rm fm^{-3}$, while it occurs near or even above $1.4 M_\odot$ with $\rho_{L}=0.4~\rm fm^{-3}$. The uncertainty of the GPR method will take the $1.6\%\sim6.4\%$ errors on the radii of NS at $2M_\odot$.  Once the uncertainty is considered, more EoSs can generate the heavier NS, whose mass is larger than $2.5 M_\odot$.

\begin{figure}[htb]
	  \centering 
	  \vspace{-5mm}  
	  \includegraphics[scale=0.3]{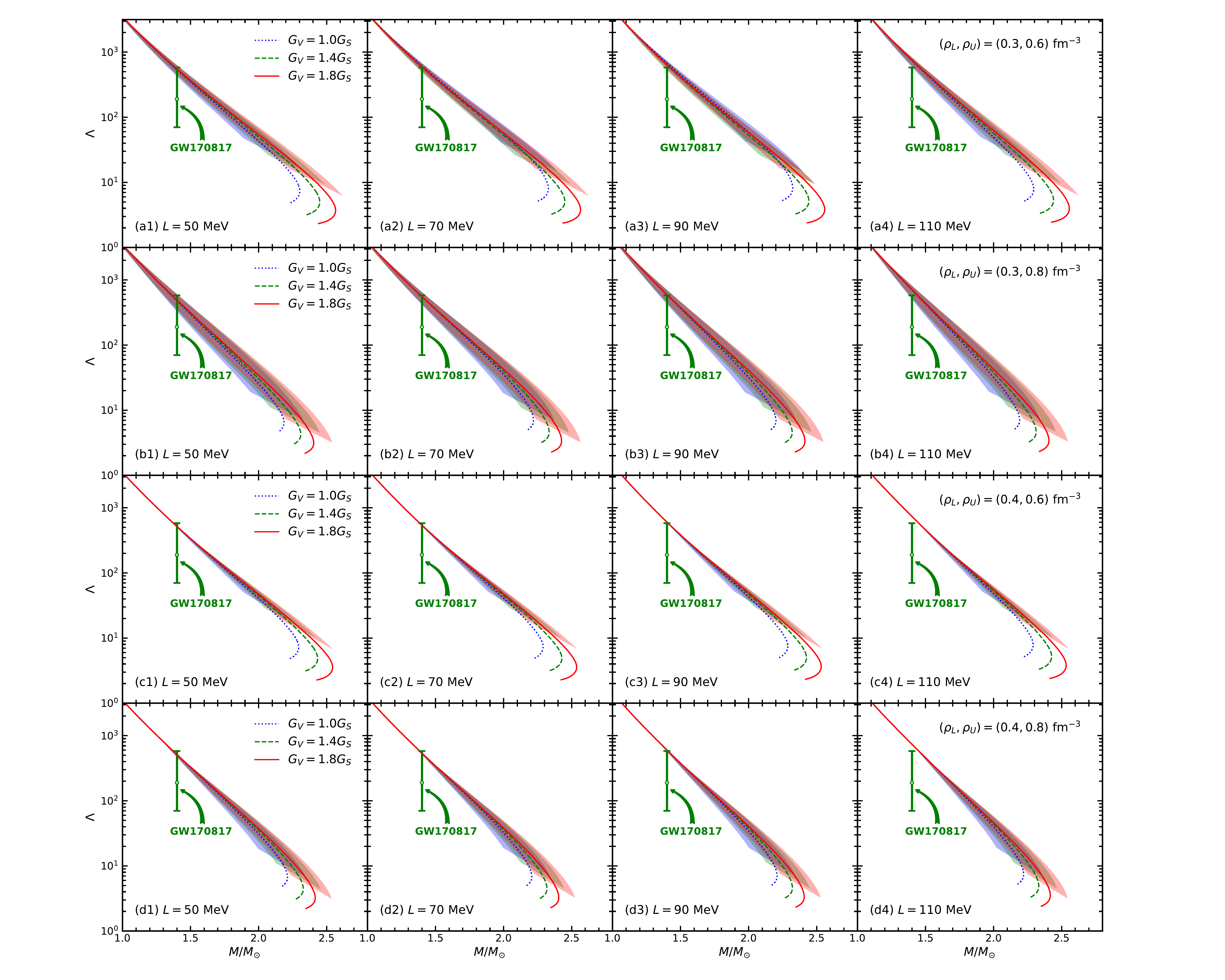}
	  \caption{The tidal deformability as a function of neutron star mass with GPR interpolation. The parameter sets in hadronic phase and quark phase, the crossover windows are the same as Fig.~\eqref{fig.GPR_P-e}.}\label{fig.GPR_lam}
\end{figure}

The dimensionless tidal deformability, $\Lambda$ of NS, is extracted by the gravitational wave detection in BNS mergers, which can denote the deformation of an NS in an external gravitational field. $\Lambda$ is related to the mass, radius of NS, with the definition $\Lambda=(2/3)k_2\left[\left(c^2/G\right)\left(R/M\right)\right]^5$, where $k_2$ is the second Love number~\citep{Hinderer2008,Hinderer2010} and $R$ is the stellar radius and $M$ the stellar mass. The gravitational wave event of BNS merger, GW170817, provides the constraint on $\Lambda$ at $M_{1.4}$ with $\Lambda_{1.4}=190^{+390}_{-120}$~\citep{Abbott2018}. In Fig.~\eqref{fig.GPR_lam},  the $\Lambda$ as function of NS mass from the EoSs with the GPR interpolation is shown. In the NS consisting of pure hadronic matter, $\Lambda$ is very sensitive to $L$ and there is a strong linear relationship between them~\citep{Hu2020}. In the present framework including the hadron-quark crossover, the dimensionless tidal deformability is dependent on not only the $L$ but also the crossover window. \replaced{The $\Lambda$ is around $540$ at $L=70$ MeV}{The $\Lambda$ is around $475-550$ at $L=50$ MeV, and around $540-550$ at $L=70$ MeV}, which satisfies the measurement of GW170817. The larger slope of symmetry energy can generate a bigger $\Lambda$, which will increase $30\%\sim40\%$ from \replaced{$L=70$ MeV}{$L=50$ MeV} to $L=110$ MeV at a same crossover window.  If the hadron-quark crossover appears earlier, the tidal deformability with the uncertainty will reduce for the light NS.
	  
Properties of neutron stars, i.e., the maximum mass ($M_{\rm max}$), the corresponding radius ($R_{\rm max}$), the central density density ($\rho_c$), the radius ($R_{1.4}$), and dimensionless tidal deformability ($\Lambda_{1.4}$) at $M_{1.4}$ with different slope of symmetry energy $L$ for hadronic phase, different vector coupling strength $G_V$ for quark phase, and various crossover windows are summarized in Table.\eqref{table.NS1} \added{and Table.\eqref{table.NS2}}. The maximum masses of NS from present EoSs are around \replaced{$2.20\sim2.55 M_\odot$}{$2.20\sim2.56 M_\odot$}. The correspond radii are \replaced{$11.14\sim 12.09$ km}{$11.00\sim 12.09$ km}. The radii at $1.4 M_\odot$ are in the range between  $12.58$ km and $13.58$ km. All of them are obviously influenced by the  crossover windows. If the mean speeds of the sound of these EoSs exceed the speed of light, they are shown in the last column.

\begin{table}[htb]
	\centering
	\tiny
	\begin{longtable}{c|c|c|cccccccccc} 
		\caption{Properties of Neutron stars obtained from the GPR interpolation between IUFSU sets for hadronic matter with $L=50,~70,~90,~110$ MeV and NJL models for quark matter with $G_V=1.0,~1.4,~1.8G_S$. The crossover windows are chosen to be  $(\rho_{L},~\rho_{U})=(0.3,~0.6),~(0.3,~0.8)~\rm fm^{-3}$.} \label{table.NS1}\\
		\hline\hline
		\multicolumn{3}{c|}{}&$M_{\rm max}/M_{\odot}$ &$R_{\rm max}~[\rm km]$ &$\rho_{\rm max}~[\rm 
		fm^{-3}]$ &$R_{\rm 1.4}~[\rm km]$ &$\rho_{\rm 1.4}~[\rm fm^{-3}]$ 
		&$\Lambda_{\rm 1.4}$ &$v_s^2$ \\
		\hline  
		\multirow{15}{2cm}{$(\rho_{L},~\rho_{U})=(0.3,~0.6)~\rm ~fm^{-3}$}
		&\multirow{3}*{\added{$L=50\rm~MeV$}}   
		&$G_V=1.0G_S$ &2.3025 &11.5354 &0.9125 &12.4540 &0.4104 &535 &\\                                                            
		&&$G_V=1.4G_S$ &2.4494 &11.7113 &0.8589 &12.4840 &0.4000 &545 & \\
		&&$G_V=1.8G_S$ &2.5659 &11.8646 &0.8151 &12.5081 &0.3920 &555 & \\
		\cline{2-10}
		&\multirow{3}*{$L=70\rm~MeV$}   
		&$G_V=1.0G_S$ &2.3291 &11.8668 &0.8687 &12.9295 &0.3764 &615 & \\                                                            
		&&$G_V=1.4G_S$ &2.4498 &11.8211 &0.8582 &12.8544 &0.3898 &568 & \\
		&&$G_V=1.8G_S$ &2.5654 &11.9639 &0.8137 &12.8752 &0.3826 &595 & \\
		\cline{2-10}
		&\multirow{3}*{$L=90\rm~MeV$}    
		&$G_V=1.0G_S$ &2.3240 &11.9454 &0.8674 &12.2580 &0.3680 &680 & \\                                 
		&&$G_V=1.4G_S$ &2.4430 &11.8782 &0.8546 &13.1560 &0.3874 &630 & \\
		&&$G_V=1.8G_S$ &2.5592 &12.0113 &0.8143 &13.1775 &0.3802 &652 & \\
		\cline{2-10}
		&\multirow{3}*{$L=110\rm~ MeV$}   
		&$G_V=1.0G_S$ &2.2950 &11.8337 &0.9001 &13.5402 &0.3805 &730 & \\
		&&$G_V=1.4G_S$ &2.4404 &11.9687 &0.8526 &13.5691 &0.3727 &749 & \\
		&&$G_V=1.8G_S$ &2.5562 &12.0912 &0.8145 &13.5852 &0.3665 &762 & \\
		\hline
		\multirow{15}{2cm}{$(\rho_{L},~\rho_{U})=(0.3,~0.8)~\rm~ fm^{-3}$} 
		&\multirow{3}*{\added{$L=50\rm~ MeV$}}   
		&$G_V=1.0G_S$ &2.1867 &10.9413 &1.0188 &12.3075 &0.4580 &475 & \\                                                            
		&&$G_V=1.4G_S$ &2.3098 &11.0035 &0.9786 &12.3300 &0.4475 &485 & \\
		&&$G_V=1.8G_S$ &2.4049 &11.0669 &0.9444 &12.3550 &0.4390 &492 &$>c^2$ \\
		\cline{2-10}
		&\multirow{3}*{$L=70\rm~ MeV$}     
		&$G_V=1.0G_S$ &2.2197 &11.2667 &0.9772 &12.7724 &0.4160 &550 &    \\
		&&$G_V=1.4G_S$ &2.3345 &11.2607 &0.9562 &12.7780 &0.4128 &550 &   \\
		&&$G_V=1.8G_S$ &2.4259 &11.2956 &0.9268 &12.7900 &0.4085 &555 &   \\
		\cline{2-10}
		&\multirow{3}*{$L=90\rm~ MeV$}     
		&$G_V=1.0G_S$ &2.2014 &11.2437 &0.9910 &13.0454 &0.4215 &587 &    \\
		&&$G_V=1.4G_S$ &2.3205 &11.2681 &0.9570 &13.0580 &0.4158 &590 &    \\
		&&$G_V=1.8G_S$ &2.4138 &11.3057 &0.9273 &13.0757 &0.4100 &600 &$>c^2$   \\
		\cline{2-10}
		&\multirow{3}*{$L=110\rm ~MeV$}    
		&$G_V=1.0G_S$ &2.1921 &11.3057 &0.9922 &13.4720 &0.4054 &705 &     \\
		&&$G_V=1.4G_S$ &2.3130 &11.3250 &0.9579 &13.4820 &0.4002 &705 &    \\
		&&$G_V=1.8G_S$ &2.4065 &11.3574 &0.9279 &13.4980 &0.3948 &712 &$>c^2$  \\ 
		\hline\hline
		
	\end{longtable}
\end{table}

\begin{table}[htb]
	\centering
	\tiny
	\begin{longtable}{c|c|c|cccccccccc} 
		\caption{Properties of Neutron stars obtained from the GPR interpolation between IUFSU sets for hadronic matter with $L=50,~70,~90,~110$ MeV and NJL models for quark matter with $G_V=1.0,~1.4,~1.8G_S$. The crossover windows are chosen to be  $(\rho_{L},~\rho_{U})=(0.4,~0.6),~(0.4,~0.8)~\rm fm^{-3}$.} \label{table.NS2}\\
		\hline\hline
		\multicolumn{3}{c|}{}&$M_{\rm max}/M_{\odot}$ &$R_{\rm max}~[\rm km]$ &$\rho_{\rm max}~[\rm 
		fm^{-3}]$ &$R_{\rm 1.4}~[\rm km]$ &$\rho_{\rm 1.4}~[\rm fm^{-3}]$ 
		&$\Lambda_{\rm 1.4}$ &$v_s^2$ \\
		\hline  
		\multirow{15}{2cm}{$(\rho_{L},~\rho_{U})=(0.4,~0.6)~\rm~fm^{-3}$} 
		&\multirow{3}*{\added{$L=50\rm~ MeV$}}   
		&$G_V=1.0G_S$ &2.2941 &11.4886 &0.9134 &12.4105 &0.4202 &515 & \\                                                            
		&&$G_V=1.4G_S$ &2.4336 &11.5908 &0.8797 &12.4125 &0.4125 &520 & \\
		&&$G_V=1.8G_S$ &2.5437 &11.6991 &0.8420 &12.4155 &0.4126 &515 &$>c^2$ \\
		\cline{2-10} 
		&\multirow{3}*{$L=70\rm~ MeV$}     
		&$G_V=1.0G_S$ &2.2903 &11.5793 &0.9141 &12.7457 &0.4152 &539 &  \\
		&&$G_V=1.4G_S$ &2.4279 &11.6816 &0.8705 &12.7485 &0.4116 &540 &   \\
		&&$G_V=1.8G_S$ &2.5368 &11.7601 &0.8422 &12.7452 &0.4094 &542 &$>c^2$  \\
		\cline{2-10}
		&\multirow{3}*{$L=90\rm~ MeV$}    
		&$G_V=1.0G_S$ &2.2880 &11.6641 &0.9137 &13.0820 &0.4062 &602 &    \\
		&&$G_V=1.4G_S$ &2.4254 &11.7575 &0.8701 &13.0835 &0.4031 &601 &     \\
		&&$G_V=1.8G_S$ &2.5338 &11.8282 &0.8416 &13.0820 &0.4012 &601 &$>c^2$ \\
		\cline{2-10}
		&\multirow{3}*{$L=110\rm ~MeV$}    
		&$G_V=1.0G_S$ &2.2906 &11.8117 &0.9006 &13.5421 &0.3858 &740 &     \\
		&&$G_V=1.4G_S$ &2.4263 &11.8738 &0.8646 &13.5421 &0.3858 &740 &     \\
		&&$G_V=1.8G_S$ &2.5333 &11.9421 &0.8324 &13.5421 &0.3858 &740 &$>c^2$ \\ 
		\hline
		\multirow{15}{2cm}{$(\rho_{L},~\rho_{U})=(0.4,~0.8)~\rm~ fm^{-3}$} 
		&\multirow{3}*{\added{$L=50\rm~ MeV$}}     
		&$G_V=1.0G_S$ &2.2126 &11.0800 &1.0061 &12.4003 &0.4328 &512 & \\                                                            
		&&$G_V=1.4G_S$ &2.3286 &11.1029 &0.9727 &12.4035 &0.4295 &510 & \\
		&&$G_V=1.8G_S$ &2.4160 &11.1244 &0.9444 &12.4054 &0.4270 &510 &$>c^2$ \\
		\cline{2-10} 
		&\multirow{3}*{$L=70\rm~ MeV$}     
		&$G_V=1.0G_S$ &2.2073 &11.1928 &0.9896 &12.7405 &0.4244 &540 &     \\
		&&$G_V=1.4G_S$ &2.3186 &11.1611 &0.9752 &12.7400 &0.4222 &538 &    \\     
		&&$G_V=1.8G_S$ &2.4053 &11.1806 &0.9417 &12.7430 &0.4205 &540 &$>c^2$   \\
		\cline{2-10}
		&\multirow{3}*{$L=90\rm ~MeV$}    
		&$G_V=1.0G_S$ &2.2097 &11.2870 &0.9908 &13.0800 &0.4138 &598 &    \\
		&&$G_V=1.4G_S$ &2.3219 &11.2814 &0.9565 &13.0800 &0.4120 &600 &       \\     
		&&$G_V=1.8G_S$ &2.4085 &11.2904 &0.9265 &13.0812 &0.4102 &600 &$>c^2$   \\
		\cline{2-10}
		&\multirow{3}*{$L=110\rm~ MeV$}   
		&$G_V=1.0G_S$ &2.2209 &11.4754 &0.9676 &13.5420 &0.3860 &738 &  \\
		&&$G_V=1.4G_S$ &2.3322 &11.4488 &0.9387 &13.5420 &0.3860 &738 &  \\     
		&&$G_V=1.8G_S$ &2.4166 &11.4388	&0.9120 &13.5420 &0.3858 &738 &$>c^2$  \\
		\hline\hline
	\end{longtable}
\end{table}	

	\section{Summaries and perspectives}\label{summary}
	The interpolated equation of state (EoS) to describe the hadron-quark crossover in neutron star (NS) was calculated with the Gaussian progress regression (GPR) method by connecting the hadronic matter at low density and quark matter at high density. The hadronic matter was described by the relativistic mean-field (RMF) model, where the isoscalar parameters were fixed and the isovector ones were readjusted to study the symmetry energy effects. The Nambu-Jona-Lasinio (NJL) model was employed to study the quark matter. A vector interaction between quarks was introduced to generate an additional repulsive contribution.
	
	The effects of the slope of symmetry energy, crossover windows, and vector interaction of the NJL model on the EoSs and properties of NS were systematically investigated. A larger slope of symmetry energy generates a bigger NS radius and larger tidal deformability. The stronger vector interaction between quarks can produce a heavier NS. The crossover window will affect the speed of sound of NS matter and the properties of NS at the inter-media mass region. The uncertainties of the interpolation method on the EoS and properties of NS were clearly shown in the GPR method in the $95\%$ credible intervals.
	
	All of these EoSs including the hadron-quark crossover can generate the massive NS, whose mass is heavier than $2.20M_\odot$. However, with the latest observables of massive NS, gravitational wave detection (GW170817 and GW190814), mass-radius of NS simultaneous measurements (NICER), \added{charged pion spectra experiment (S$\pi$RIT )}, and the neutron skin  thickness of $^{208}{\rm Pb}$ (PREX-II), the slope of symmetry energy was predicted around \replaced{$70-90$ MeV}{$50-90$ MeV}, the crossover window was expected as $(0.3,~0.6)~\rm fm^{-3}$, and the vector coupling strength $G_V$ should be less than $1.8G_S$ in the present framework.
	
	 There are many different schemes to discuss the existence of quarks in the core of NS. However, it is still a puzzle about the hadron-quark phase transition process. The crossover picture avoids the complicated mechanism of phase transition and provides a way to generate a massive NS. It is very hard to estimate the uncertainties of results and has arbitrariness in the previous interpolation methods. The GPR method as a nonparametric Bayesian approach can calculate the uncertainty of the predictions with the training data. Therefore, the EoSs are generated by the GPR method and present conclusions are independent of interpolation methods. The python code for the EoSs of hadron-quark crossover with the GPR method can be found in the supplementary material.
	  	      
\section{Acknowledgments}
This work was supported in part by the National Natural Science Foundation of China (Grant  Nos. 11775119 and 12175109), and  the Natural Science Foundation of Tianjin (Grant  No. 19JCYBJC30800).

\listofchanges 
\end{document}